\documentclass[aps,twocolumn,floatfix,bibnotes]{revtex4}

\usepackage{here}

\usepackage[dvips]{graphicx}

\newcommand\pictc[5]{\begin{figure}
                       \centerline{
                       \includegraphics[width=#1\columnwidth]{#3}}
                   \protect\caption{\protect\label{fig:#4} #5}
                    \end{figure}            }
\newcommand\pict[4][.7]{\pictc{#1}{!tb}{#2}{#3}{#4}}
\newcommand\rpict[1]{\ref{fig:#1}}

\newcommand\leqt[1]{\protect\label{eq:#1}}

\newcounter{Fig}

\begin{document}

\begin{sloppy}

\title{Nonlinear left-handed metamaterials}

\author{Ilya V. Shadrivov$^1$}
\author{Alexander A. Zharov$^{1,2}$}
\author{Nina A. Zharova$^{1,3}$}
\author{Yuri S. Kivshar$^1$}

\affiliation{$^1$Nonlinear Physics Centre, Research School of
Physical Sciences and Engineering, Australian National University,
Canberra ACT 0200, Australia \\
$^2$ Institute for Physics of Microstructures, Russian Academy of
Sciences, Nizhny Novgorod 603950, Russia\\
$^3$ Institute of Applied Physics, Russian Academy of Sciences, Nizhny
 Novgorod 603600, Russia}

\begin{abstract}
We analyze nonlinear properties of microstructured materials with
the negative refractive index, the so-called {\em left-handed
metamaterials}. We demonstrate that the hysteresis-type dependence
of the magnetic permeability on the field intensity allows
changing the material properties from left- to right-handed and
back. Using the finite-difference time-domain simulations, we
study the wave reflection from a slab of nonlinear
left-handed material, and observe the propagation of {\em temporal
solitons} in such materials. We demonstrate also that nonlinear
left-handed metamaterials can support both TE- and TM-polarized
self-trapped localized beams, {\em spatial electromagnetic
solitons}. Such solitons appear as single- and multi-hump beams,
being either symmetric or antisymmetric, and they can exist due to
the hysteresis-type magnetic nonlinearity and the effective
domains of negative magnetic permeability.
\end{abstract}

\maketitle

\section{Introduction}

Recent theoretical studies
\cite{Pendry:1996-4773:PRL,Pendry:1999-2075:ITMT,markos} and
experimental results \cite{Smith:2000-4184:PRL,bay,prl_new} have
shown the possibility of creating novel types of microstructured
materials that demonstrate many intriguing properties such as the
effect of negative refraction. In particular, the composite
materials created by arrays of wires and split-ring resonators
were shown to possess a negative real part of the magnetic
permeability and dielectric permittivity for microwaves. These
materials are often referred to as {\em left-handed materials}
(LHMs) or {\em materials with negative refraction}. Properties of
the left-handed materials were analyzed theoretically by Veselago
long time ago~\cite{veselago}, but such materials were
demonstrated experimentally only very recently. As was shown by
Veselago~\cite{veselago}, the left-handed materials possess a
number of peculiar properties, including negative refraction for
interface scattering, inverse light pressure, reverse Doppler and
Vavilov-Cherenkov effects, etc.

So far, most of the properties of left-handed materials were
studied in the linear regime of wave propagation when both
magnetic permeability and dielectric permittivity of the material
are assumed to be independent on the intensity of the
electromagnetic field. However, any future effort in creating {\em
tunable structures} where the field intensity changes the
transmission properties of the composite structure would require
the study of nonlinear properties of such metamaterials, which may
be quite unusual. In particular, the recently fabricated
metamaterials are composed of a mesh of wires and split-ring
resonators (SRRs). The wires provides negative dielectric
permittivity, while SRRs give negative magnetic permeability.
Metamaterials possess left-handed properties only in some finite
frequency range, which is basically determined by the geometry of
the structure. The possibility to control the effective parameters
of the metamaterial using nonlinearity has recently been suggested
in Refs.~\cite{our_prl,lapine}. Importantly, the microscopic
electric field in the left-handed structure can be much higher
than the macroscopic electric field carried by the propagating
wave. This provides a simple physical mechanism for enhancing
nonlinear effects in left-handed materials.

In this paper we present a brief overview of nonlinear properties
of left-handed metamaterials for the example of a lattice of the
split-ring resonators and wires with a nonlinear dielectric. By
means of the finite-difference time-domain (FDTD) simulations, we
study the wave scattering on a slab of a nonlinear composite
structure. We also discuss the structure of electromagnetic
solitons supported by the nonlinear left-handed materials with
hysteresis-type nonlinear response. We believe our findings may
stimulate the future experiments in this field, as well as the
studies of nonlinear effects in photonic crystals, where the
phenomenon of negative refraction is analyzed now very
intensively~\cite{pbg,pbg1}.

\section{Nonlinear resonant response}

First, we follow the original paper~\cite{our_prl} and consider a
two-dimensional composite structure consisting of a square lattice
of the periodic arrays of conducting wires and split-ring
resonators (SRR). We assume that the unit-cell size $d$ of the
structure is much smaller then the wavelength of the propagating
electromagnetic field and, for simplicity,  we choose the
single-ring geometry of a lattice of cylindrical SRRs. The results
obtained for this case are qualitatively similar to those obtained
in the more involved cases of double SRRs. This type of
microstructured materials has recently been suggested and built in
order to create left-handed metamaterials with negative refraction
in the microwave region~\cite{Smith:2000-4184:PRL}.

The negative real part of the effective dielectric permittivity of
such a composite structure appears due to the metallic wires
whereas a negative sign of the magnetic permeability becomes
possible due to the SRR lattice. As a result, these materials
demonstrate the properties of negative refraction in the finite
frequency band, $\omega_{0} < \omega < {\rm min} (\omega_p,
\omega_{\parallel m})$, where $\omega_{0}$ is the eigenfrequency
of the SRRs, $\omega_{\parallel m}$ is the frequency of the
longitudinal magnetic plasmon, $\omega_p$ is an effective plasma
frequency, and $\omega$ is the angular frequency of the
propagating electromagnetic wave, $({\cal E},{\cal H})\sim ({\bf
E},{\bf H})\exp{(i\omega t)}$. The split-ring resonator can be
described as an effective LC oscillator (see, e.g. Ref.
\cite{Gorkunov:2002-263:EPB}) with the capacitance of the SRR gap,
as well as an effective inductance and resistance.

Nonlinear response of such a composite structure can be
characterized by two different contributions. The first one is an
intensity-dependent part of the effective dielectric permittivity
of the infilling dielectric. For simplicity, we may assume that
the metallic structure is embedded into a nonlinear dielectric
with a permittivity that depends on the intensity of the electric
field in a general form,  $ \epsilon_D = \epsilon_D(|{\bf E}|^2)$.
For results of calculations presented below, we take the linear
dependence that corresponds to the Kerr-type nonlinear response.

The second contribution into the nonlinear properties of the
composite material comes from the lattice of resonators, since the
SRR capacitance (and, therefore, the SRR eigenfrequency) depends
on the strength of the local electric field in a narrow slot.
The intensity of the local electric field in the SRR gap
depends on the electromotive force in the resonator loop, which is
induced by the magnetic field. Therefore, the effective magnetic
permeability $\mu_{\rm eff}$ depends on the macroscopic (average)
magnetic field ${\bf H}$, and this dependence can be found in the
form~\cite{our_prl}
\begin{equation}
\label{mu_eff}
\mu_{\rm eff}({\bf H}) = 1 + \frac{F\,
\omega^2}{\omega_{0NL}^2({\bf H}) -
                \omega^2 + i \Gamma \omega},
\end{equation}
where
\[
\omega_{0NL}^2({\bf H})= \left(\frac{c}{a}\right)^2
\frac{d_g}{[\pi h \epsilon_D(|{\bf E}_g({\bf H})|^2)]}
\]
is the eigenfrequency of oscillations in the presence of the
external field of a finite amplitude, $h$ is the width of the
ring, $\Gamma=c^2/2\pi\sigma ah$,  for $h<\delta$,  and
$\Gamma=c^2/2\pi\sigma a\delta$, for $h>\delta$. It is important
to note that Eq.~(\ref{mu_eff}) has a simple physical
interpretation: The resonant frequency of the artificial magnetic
structure depends on the amplitude of the external magnetic field
and, in turn, this leads to the intensity-dependent function
$\mu_{\rm eff}$.

Figures~\rpict{Re_mu} and \rpict{Im_mu} summarize different types
of nonlinear magnetic properties of the composite, which are
defined by the dimensionless frequency of the external field
$\Omega=\omega/\omega_0$, for both a {\em focusing}
[Figs.~\rpict{Re_mu}, \rpict{Im_mu}(a,b)] and a {\em defocusing}
[Figs.~\rpict{Re_mu}, \rpict{Im_mu}(c,d)] nonlinearity of the
dielectric.
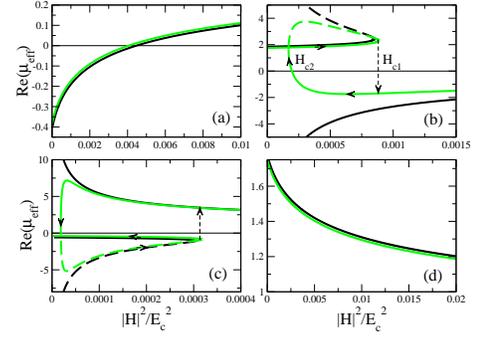
\pict{fig01.eps}{Re_mu}{Real part of the effective magnetic permeability vs.
intensity of the magnetic field: (a)~$\Omega>1$, $\alpha=1$;
(b)~$\Omega<1$, $\alpha=1$, (c)~$\Omega>1$, $\alpha=-1$; and
(d)~$\Omega<1$, $\alpha=-1$. Black -- the lossless case ($\gamma =
0$), grey--the lossy case ($\gamma = 0.05$). Dashed curves show
unstable branches.}
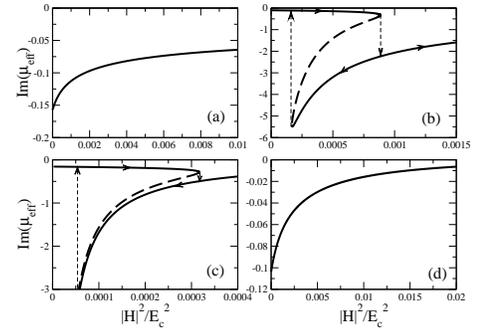
\pict{fig02.eps}{Im_mu}{Imaginary part of the effective magnetic
permeability vs. intensity of the magnetic field for $\gamma =
0.05$: (a)~$\Omega>1$, $\alpha=1$; (b)~$\Omega<1$, $\alpha=1$,
(c)~$\Omega>1$, $\alpha=-1$; and (d)~$\Omega<1$, $\alpha=-1$.
Dashed curves show unstable branches.}

Due to the high values of the electric field in the slot of SRR as
well as resonant interaction of the electromagnetic field with the
SRR lattice, the characteristic magnetic nonlinearity in such
structures is much stronger then the corresponding electric
nonlinearity. Therefore, {\em magnetic nonlinearity should
dominate} in the composite metamaterials. More importantly, the
nonlinear medium can be created by inserting nonlinear elements
into the slots of SRRs, allowing an easy tuning by an external
field.

The critical fields for switching between LH and RH states, shown
in the Figs.~\rpict{Re_mu} can be reduced to a desirable value by
choosing the frequency close to the resonant frequency of SRRs.
Even for a relatively large difference between the SRR
eigenfrequency and the external frequency, as we have in
Fig.~\rpict{Re_mu}(b) where $\Omega = 0.8$ (i.e.
$\omega=0.8\omega_0$), the switching amplitude of the magnetic
field is $\sim 0.03 E_c$. The characteristic values of the
focusing nonlinearity can be estimated for some materials such as
n-InSb for which $E_c = 200 V/cm$~\cite{Belyantsev}. As a result,
the strength of the critical magnetic field is found as $H_{c1}
\approx 1.6 A/m$. Strong defocusing properties for microwave
frequencies are found in Ba$_x$Sr$_{1-x}$TiO$_3$ (see
Ref.~\cite{Li:2001-2354:APL} and references therein). The critical
nonlinear field of a thin film of this material is $E_c=4\cdot10^4
V/cm$, and the corresponding field of the transition from the LH
to RH state [see Fig.~\rpict{Re_mu} (c)] can be found as $H_{c}
\approx 55.4 A/m$.

The unique possibility of strongly enhanced effective
nonlinearities in the left-handed metamaterials revealed here may
lead to an essential revision of the concepts based on the linear
theory, since the electromagnetic waves propagating in such
materials always have a finite amplitude.  At the same time, the
engineering of nonlinear composite materials may open a number of
their novel applications such as frequency multipliers, beam
spatial spectrum transformers, switchers, limiters, etc.

\section{FDTD simulations of nonlinear transmission}

In order to verify the specific features of the left-handed
metamaterials introduced by their nonlinear response, in this
section we study the scattering of electromagnetic waves from the
nonlinear metamaterial discussed above. In particular, we perform
the FDTD numerical simulations of the plane wave interaction with
a slab of LHM of a finite thickness. We use the Maxwell's
equations in the form
\begin{eqnarray}\leqt{maxw}
\nabla \times {\bf E} = -\frac{1}{c}\frac{\partial {\bf B}}{\partial t},\nonumber\\
\nabla \times {\bf B} = \frac{1}{c}\frac{\partial {\bf E}}{\partial t} +
    \frac{4\pi}{c} \left<{\bf j}\right> +
    4\pi \nabla \times {\bf M},
\end{eqnarray}
where $\left<{\bf j}\right>$ is the current density averaged over
the period of the unit cell, and ${\bf M}$ is the magnetization of
the metamaterial. We base our analysis on the microscopic model
recently discussed in Ref.~\cite{Shadrivov:Periodic}, and write
the constitutive relations in the form
\begin{eqnarray}\leqt{mat}
\sigma L_w S \frac{d \left<{\bf j}\right>}{dt} + \left<{\bf
j}\right>=
    \frac{\sigma S}{d_{cell}^2}{\bf E},\nonumber\\
{\bf M} = \frac{n_m}{2c}\pi a^2 I_R \frac{{\bf B}}{|{\bf B}|},
\end{eqnarray}
where $L_w$ is the inductance of the wire per unit length,
$\sigma$ is the conductivity of metal in the composite, $S$ is the
effective cross-section of a wire, $S \approx \pi r_w^2$, for
$\delta > r_w$, and $S \approx \pi \delta(2r-\delta)$, for $\delta
<r_w$, where $\delta = c/\sqrt{2\pi\sigma\omega}$ is the
skin-layer thickness, $I_R$ is the current in SRR, $n_m$ is
concentration of SRRs. The current in SRRs is governed by the
equation
\begin{equation}\leqt{current}
L\frac{dI_R}{dt} = -\frac{\pi a^2}{c}\frac{dH^{\prime}}{dt} -
        U - R I_R,
\end{equation}
where $L$ is inductance of the SRR, $R$ is resistance of the SRR
wire, $U$ is the voltage on the SRR slit, and $H^{\prime}$ is the
acting (microscopic) magnetic field, which differs from the
average (macroscopic) magnetic field. Voltage $U$ at the slit of
SRR is coupled to the current $I$ through the relation
\begin{equation}\leqt{volt}
C(U)\frac{dU}{dt}=I_R,
\end{equation}
with\[C(U) = \pi r^2 \epsilon\left( 1 + \alpha
\left|U\right|^2/U_c^2 \right) / 4\pi d_g,\] where $\epsilon$ is
the linear part of the permittivity of a dielectric material
inside the SRR slit, $U_c$ is the characteristic nonlinear
voltage, and $\alpha = \pm 1$ corresponds to the case of the
focusing and defocusing nonlinear response, respectively.

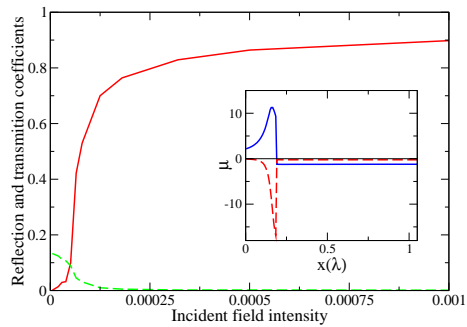
\pict{fig03.eps}{refl_intens_rough}{Reflection (solid) and
transmission (dashed) coefficients for a slab of nonlinear
metamaterial {\em vs.} the incident field intensity in a
stationary regime, for the case of defocusing nonlinearity
($\alpha = -1$). Inset shows real (solid) and imaginary (dashed)
parts of the magnetic permeability inside the slab.}

The microscopic magnetic field ${\bf H}^{\prime}$ can be expressed
in terms of ${\bf M}$ and ${\bf B}$ using the Lorenz-Lorentz
relation~\cite{Born:2002:PrinciplesOptics}:
\begin{equation}\leqt{Lorentz}
{\bf H}^{\prime} = {\bf B} -\frac{8\pi}{3}{\bf M}.
\end{equation}
As a result, Eqs.~(\ref{eq:maxw}) to (\ref{eq:Lorentz}) form a
closed system of coupled equations, and they can be solved
numerically using, for example, the numerical FDTD method. We also
notice that, by substituting the harmonic fields into these
equations, we recover the expression for the magnetic permeability
(\ref{mu_eff}).

Our goal is to study the temporal dynamics of the wave scattering
by a finite slab of nonlinear metamaterial. For simplicity, we
consider a one-dimensional problem that describes the interaction
of the plane wave incident at the normal angle from air on a slab
of metamaterial of a finite thickness. We consider {\em two types
of nonlinear effects}: (i) nonlinearity-induced suppression of the
wave transmission when initially transparent left-handed material
becomes opaque with the growth of the input amplitude, and (ii)
nonlinearity-induced transparency when an opaque metamaterial
becomes left-handed (and therefore transparent) with the growth of
the input amplitude. The first case corresponds to the dependence
of the effective magnetic permeability on the external field shown
in Figs.~\rpict{Re_mu}(a,c), when initially negative magnetic
permeability (we consider $\epsilon<0$ in all frequency range)
becomes positive with the growth of the magnetic field intensity.
The second case corresponds to the dependence of the magnetic
permeability on the external field shown in
Figs.~\rpict{Re_mu}(b).

In all numerical simulations, we use {\em linearly growing}
amplitude of the incident field within the first 50 periods, that
becomes constant afterwards. The slab thickness is selected as
$1.3\lambda_0$ where $\lambda_0$ is a free-space wavelength. For
the parameters we have chosen, the metamaterial is left-handed in the linear
regime for the frequency range from $f_1=5.787$ GHz to $f_2 =
6.05$ GHz.

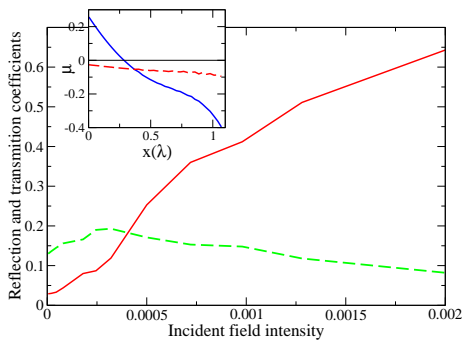
\pict{fig04.eps}{refl_intens_smoth}{Reflection (solid) and
transmission (dashed) coefficients for a slab of nonlinear
metamaterial {\em vs.} the incident field intensity in a
stationary regime, for the focusing nonlinearity ($\alpha = 1$).
Inset shows real (solid) and imaginary (dashed) parts of the
magnetic permeability inside the slab.}

Our simulations show that for the incident wave with the frequency
$f_0 = 5.9$ GHz (i.e. inside the left-handed transmission band),
electromagnetic field reaches a steady state independently of the
sign of the nonlinearity. Both reflection and transmission
coefficients  in the {\em stationary} regime are shown in
Figs.~\rpict{refl_intens_rough},\rpict{refl_intens_smoth} as
functions of the incident field amplitude, for both defocusing and
focussing nonlinear response of the dielectric infilling the SRR
slits. In the linear regime, the effective parameters of the
metamaterial at the frequency $f_0$ are: $\epsilon = -1.33-0.01i$
and $\mu = -1.27 -0.3i$; this allows excellent impedance matching
with surrounding air. The scattering results in a vanishing
reflection coefficient for small incident intensities (see
Figs.~\rpict{refl_intens_rough},\rpict{refl_intens_smoth}).

Reflection and transmission coefficients are qualitatively
different for two different types of infilling nonlinear
dielectric. For the defocusing nonlinearity, the reflection
coefficient varies from low to high values when the incident field
exceeds some threshold value (see Fig.~\rpict{refl_intens_rough}).
Such a sharp transition can be explained in terms of the
hysteresis behavior of the magnetic permeability shown in
Fig.~\rpict{Re_mu}(c). When the field amplitude in metamaterial
becomes higher than the critical amplitude [shown by a dashed
arrow in Fig.~\rpict{Re_mu}(c)], magnetic permeability changes its
sign, and the metamaterial becomes opaque. Our FDTD simulations
show that for overcritical amplitudes of the incident field, the
opaque region of positive magnetic permeability appears inside the
slab (see the inset in Fig.~\rpict{refl_intens_rough}). The
magnetic permeability experiences an abrupt change at the boundary
between the transparent and opaque regions. The dependencies shown
in Fig.~\rpict{refl_intens_rough} are obtained for the case when
the incident field grows from zero to a steady-state value.
However, taking different temporal behavior of the incident wave,
e.g. increasing the amplitude above the threshold value and then
decreasing it to the steady state, one can get different values of
the stationary reflection and transmission coefficients, and
different distributions of the magnetic permeability inside the
metamaterial slab. Such properties of the nonlinear metamaterial
slab are consistent with the predicted multi-valued dependence of
the magnetic permeability on the amplitude of the magnetic field.

In the case of focussing nonlinearity (see
Fig.~\rpict{refl_intens_smoth}), the dependence of the reflection
and transmission coefficients on the amplitude of the incident
field is smooth. This effect originates, firstly, from a gradual
detuning from the impedance matching condition, and, for higher
powers, from the appearance of an opaque layer (see the inset in
Fig.~\rpict{refl_intens_smoth}) with a positive value of the
magnetic permeability that is a continuous function of the
coordinate inside the slab.

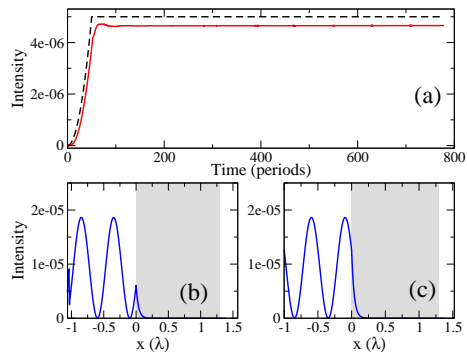
\pict{fig05.eps}{refl_lin_801}{(a) Reflected (solid) and
incident (dashed) wave intensity {\em vs} time for small
amplitudes of the incident wave (i.e. in the linear regime). (b,c)
Distribution of the magnetic and electric fields  at the end of
simulation time; the metamaterial is shaded.}
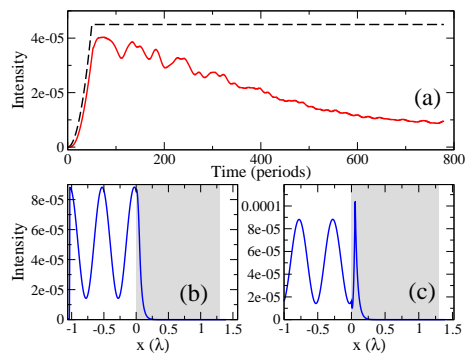
\pict{fig06.eps}{refl_int_803}{The same as in Fig.~5 but in the
regime of overcritical  amplitude of the incident wave. }

Now we consider another interesting case when initially opaque
metamaterial becomes transparent with the growth of the incident
field amplitude. We take the frequency of the incident field to be
$f_0=5.67$ GHz, so that  magnetic permeability is positive in the
linear regime and the metamaterial is opaque. In the case of
self-focussing nonlinear response ($\alpha=1$), it is possible to
switch the material properties to the regime with negative
magnetic permeability [see Fig.\rpict{Re_mu}(b) making the
material slab left-handed and therefore transparent. Moreover, one
can expect the formation of self-focused localized states inside
the composite, the effect which was previously discussed for the
interaction of the intense electromagnetic waves with over-dense
plasma~\cite{plasma}. Figure \rpict{refl_lin_801}(a) shows the
temporal evolution of the incident and reflected wave intensities
for small input intensities, this case corresponds to the linear
regime. The reflection coefficient reaches a steady state after
approximately 100 periods. The spatial distribution of the
electric and magnetic fields at the end of simulation time is
shown in Fig.~\rpict{refl_lin_801}(b,c), respectively.

In a weakly nonlinear overcritical regime (see
Fig.~\rpict{refl_int_803}), the intensity of the reflected beam
decreases approaching a steady state. In this case we observe the
formation of a localized state inside the metamaterial slab and
near the interface, as can be seen more distinctly in
Fig.~\rpict{refl_int_803}(c). This effect give an additional
contribution to the absorption of the electromagnetic energy, thus
leading to a decay of the value of the reflection coefficient.

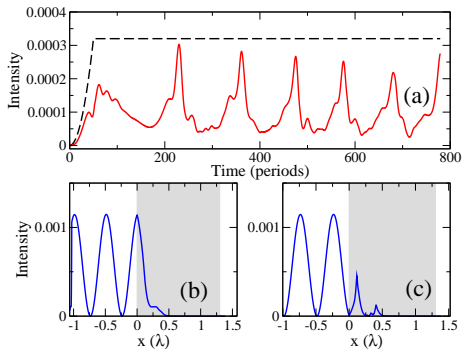
\pict{fig07.eps}{refl_nl_806}{The same as in Fig.~5 but in the
regime of strongly overcritical amplitude of the incident wave.}

In a strongly nonlinear overcritical regime, we observe the effect
of the dynamical self-modulation of the reflected electromagnetic
wave that results from the periodic generation of the
self-localized states inside the metamaterial (see
Fig.~\rpict{refl_nl_806}). Such localized states resemble {\em
temporal solitons}, which transfer the energy away from the
interface. Figure~\rpict{refl_nl_806}(c) shows an example when two
localized states enter the metamaterial. These localized states
appear on the jumps of the magnetic permeability and, as a result,
we observe a change of the sign of the electric field derivative
at the maximum of the soliton intensity, and subsequent appearance
of transparent regions in the metamaterial. Unlike all previous
cases, the field structure in this regime do not reach any steady
state for high enough intensities of the incident field.

\section{Electromagnetic spatial solitons}

Similar to other nonlinear
media~\cite{Kivshar:2003:OpticalSolitons}, nonlinear left-handed
composite materials can support self-trapped electromagnetic waves
in the form of {\em spatial solitons}. Such solitons possess
interesting properties because they exist in materials with a
hysteresis-type (multi-stable) nonlinear magnetic response. Below,
we describe novel and unique types of single- and multi-hump
(symmetric, antisymmetric, or even asymmetric) backward-wave
spatial electromagnetic solitons supported by the nonlinear
magnetic permeability.

Spatially localized TM-polarized waves that are described by one
component of the magnetic field and two components of the electric
field. Monochromatic stationary waves with the magnetic field
component $H = H_y$ propagating along the $z$-axis and homogeneous
in the $y$-direction, $[\sim \exp{(i\omega t - i k z)}]$, are
described by the dimensionless nonlinear Helmholtz equation
\begin{equation}
\label{H_field}
\frac{d^2 H}{d x^2} + [\epsilon \mu(|H|^2) -
\gamma^2]H = 0,
\end{equation}
where $\gamma = k c/\omega$ is a wavenumber, $x =
x^{\prime}\omega/c$ is the dimensionless coordinate, and
$x^{\prime}$ is the dimensional coordinate. Different types of
localized solutions of Eq.~(\ref{H_field}) can be analyzed on the
phase plane $(H, dH/dx)$ (see, e.g.,
Refs.~\cite{Gildenburg:1983-48:ZETF}). First, we find the
equilibrium points: the point $(0,0)$ existing for all parameters,
and the point $(0,H_1)$, where $H_1$ is found as a solution of the
equation
\begin{equation}
\label{X_eq}
X^2(H_1) = X^2_{\rm eq} = \Omega^2 \left\{1 +
\frac{F\epsilon_{\rm eff}} {(\gamma^2 - \epsilon_{\rm eff})}
\right\}.
\end{equation}
Below the threshold, i.e. for $\gamma < \gamma_{\rm tr}$, where
$\gamma_{\rm tr}^2 = \epsilon [1+F\Omega^2/(1-\Omega^2)]$, the
only equilibrium state $(0,0)$ is a saddle point and, therefore,
no finite-amplitude or localized waves can exist. Above the
threshold value, i.e. for $\gamma > \gamma_{\rm tr}$, the phase
plane has three equilibrium points, and a separatrix curve
corresponds to a soliton solution.
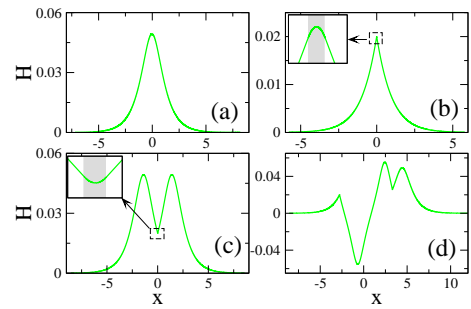
\pict{fig08.eps}{solitons}{Examples of different types of solitons:
(a) fundamental soliton; (b,c) solitons with one domain of
negative or positive magnetic permeability (shaded), respectively;
(d) soliton with two different domains (shaded). Insets in (b,c)
show the magnified regions of the steep change of the magnetic
field.}

In the vicinity of the equilibrium state $(0,0)$, linear solutions
of Eq.~(\ref{H_field}) describe either exponentially growing or
exponentially decaying modes. The equilibrium state $(0, H_1)$
describes a finite-amplitude wave mode of the transverse
electromagnetic field. In the region of multi-stability, the type
of the phase trajectories is defined by the corresponding branch
of the multi-valued magnetic permeability. Correspondingly,
different types of the spatial solitons appear when the phase
trajectories correspond to the different branches of the nonlinear
magnetic permeability.

The fundamental soliton is described by the separatrix trajectory
on the plane $(H, dH/dx)$ that starts at the point $(0, 0)$, goes
around the center point $(0, H_1)$, and then returns back; the
corresponding soliton profile is shown in Fig.~\rpict{solitons}(a).
More complex solitons are formed when the magnetic permeability
becomes multi-valued and is described by several branches. Then,
soliton solutions are obtained by switching between the separatrix
trajectories corresponding to different (upper and lower) branches
of magnetic permeability. Continuity of the tangential components
of the electric and magnetic fields at the boundaries of the
domains with different values of magnetic permeability implies
that both $H$ and $dH/dx$ should be continuous. As a result, the
transitions between different phase trajectories should be
continuous.

Figures~\rpict{solitons}(b,c) show several examples of the more
complex solitons corresponding to a single jump to the lower
branch of $\mu(H)$ (dotted) and to the upper branch of $\mu(H)$
(dashed), respectively. The insets show the magnified domains of a
steep change of the magnetic field. Both the magnetic field and
its derivative, proportional to the tangential component of the
electric field, are continuous. The shaded areas show the
effective domains where the value of magnetic permeability
changes. Figure~\rpict{solitons}(d) shows an example of more
complicated multi-hump soliton which includes two domains of the
effective magnetic permeability, one described by the lower
branch, and the other one -- by the upper branch. In a similar
way, we can find more complicated solitons with different number
of domains of the effective magnetic permeability.

We note that some of the phase trajectories have discontinuity of
the derivative at $H=0$ caused by infinite values of the magnetic
permeability at the corresponding branch of $\mu_{\rm eff}(H)$.
Such a non-physical effect is an artifact of the lossless model of
a left-handed nonlinear composite considered here for the analysis
of the soliton solutions. In more realistic models that include
losses, the region of multi-stability does not extend to the point
$H = 0$, and in this limit the magnetic permeability remains a
single-valued function of the magnetic field~\cite{our_prl}.

For such a multi-valued nonlinear magnetic response, the domains
with different values of the magnetic permeability "excited" by
the spatial soliton can be viewed as effective induced left-handed
waveguides which make possible the existence of single- and
multi-hump soliton structures. Due to the existence of such
domains, the solitons can be not only symmetric, but also
antisymmetric and even asymmetric. Formally, the size of an
effective domain can be much smaller than the wavelength and,
therefore, there exists an applicability limit for the obtained
results to describe nonlinear waves in realistic composite
structures.

When the infilling dielectric of the structure displays {\em
self-focusing nonlinear response}, we have $\Omega <1$, and in
such system we can find {\em dark solitons}, i.e. localized dips
on the finite-amplitude background
wave~\cite{Kivshar:2003:OpticalSolitons}.  Similar to bright
solitons, there exist both fundamental dark solitons and dark
solitons with domains of different values of magnetic
permeability. For self-defocusing nonlinearity and $\Omega<1$,
magnetic permeability is a single-valued function, and such a
nonlinear response can support dark solitons as well, whereas for
self-focusing dielectric, we have $\Omega> 1$ and no dark solitons
can exist.

\section{Conclusions}

We have discussed novel properties of left-handed metamaterials
associated with their  nonlinear resonant response. For the case
of harmonic fields, we have calculated the effective magnetic
permeability of microstructured materials consisting of rods and
split-ring resonators, and predicted the hysteresis-like
dependence of the nonlinear magnetic permeability as a function of
the applied magnetic field. Using the finite-difference
time-domain numerical simulations, we have studied the temporal
dynamics of the wave reflection from a slab of nonlinear
metamaterial that is found to be consistent with our theory.
Finally, we have predicted the existence of electromagnetic
spatial solitons supported by the hysteresis-type nonlinear
magnetic permeability of a left-handed material.

The work has been supported by the Australian Research Council and
the Australian National University.

\end{sloppy}
\end{document}